# SCIENTIFIC REPORTS

**OPEN**

# Root Mean Square Minimum Distance as a Quality Metric for Stochastic Optical Localization Nanoscopy Images

Yi Sun

A localization algorithm in stochastic optical localization nanoscopy plays an important role in obtaining a high-quality image. A universal and objective metric is crucial and necessary to evaluate qualities of nanoscopy images and performances of localization algorithms. In this paper, we propose root mean square minimum distance (RMSMD) as a quality metric for localization nanoscopy images. RMSMD measures an average, local, and mutual fitness between two sets of points. Its properties common to a distance metric as well as unique to itself are presented. The ambiguity, discontinuity, and inappropriateness of the metrics of accuracy, precision, recall, and Jaccard index, which are currently used in the literature, are analyzed. A numerical example demonstrates the advantages of RMSMD over the four existing metrics that fail to distinguish qualities of different nanoscopy images in certain conditions. The unbiased Gaussian estimator that achieves the Fisher information and Cramer-Rao lower bound (CRLB) of a single data frame is proposed to benchmark the quality of localization nanoscopy images and the performance of localization algorithms. The information-achieving estimator is simulated in an example and the result demonstrates the superior sensitivity of RMSMD over the other four metrics. As a universal and objective metric, RMSMD can be broadly employed in various applications to measure the mutual fitness of two sets of points.

A localization algorithm or an estimator that estimates emitter locations plays an important role in obtaining a high-quality image in stochastic optical localization nanoscopy. A number of localization algorithms and software packages have been developed in the literature. It is imperative to identify the good algorithms, which yield a high quality of nanoscopy images, and to advance research and development of localization algorithms. To achieve this goal, by challenging 2D synthetic nanoscopy data, performances of thirty two localization software packages were recently evaluated[1]. The challenge has been advanced to focus on 3D imaging[2] and become an open public online challenge that has drawn eighty four participant packages[3]. A quality metric for localization nanoscopy images is crucial and necessary in research and development of localization algorithms. However, a localization nanoscopy image is substantially different from a conventional image. The former consists of a set of points defined over an $n$-dimensional real space while the latter consists of an array of pixels with a finite number of intensities. A quality metric for the conventional images usually fails to evaluate the quality of the localization nanoscopy images. To emulate a quality metric for the conventional images, Fourier ring correlation (FRC) has been proposed to evaluate resolution of localization nanoscopy images[4,5]. Cramer Rao lower bound (CRLB)[6,7] can be employed to evaluate the minimum variance between an estimated emitter location and its true location for all unbiased localization algorithms. Nevertheless, these metrics are unable to manifest the quality of a localization nanoscopy image as an estimate of a set of emitter locations. A universal and objective metric, which is crucial and necessary to evaluate qualities of nanoscopy images and performances of localization algorithms, has not yet been established in the field[2]. In the current research, accuracy, precision, recall, and Jaccard index (JAC) are used as quality metrics for nanoscopy images[1,3,8,9]. These metrics depend on the full-width half-maximum (FWHM) of the point spread function (PSF) in an optical system and therefore are not universal and somewhat subjective, and in certain conditions fail to distinguish qualities of different nanoscopy images.

Electrical Engineering Department, Nanoscopy Laboratory, The City College of City University of New York, New York, NY, 10031, USA. Correspondence and requests for materials should be addressed to Y.S. (email: ysun@ccny.cuny.edu)

In this paper, we propose root mean square minimum distance (RMSMD) as a quality metric for a localization nanoscopy image when the set of emitter locations is known. RMSMD is defined as root mean square of minimum distances from a point of one set to another set and vice versus. RMSMD measures the average, local, and mutual fitness between two sets of points. The smaller the RMSMD, the better the fitness or the higher quality of one set in description of another. The Voronoi cell of a point can be employed to express RMSMD. RMSMD has several properties that are common to a distance metric as RMSMD is. The kernel points of the two sets play an important role in RMSMD since removal of non-kernel points can usually reduce RMSMD. A numerical example demonstrates that RMSMD can properly measure the quality of a sequence of localization nanoscopy images while the metrics of accuracy, precision, recall, and JAC fail to distinguish the qualities of different nanoscopy images in certain conditions. The ambiguity, discontinuity, and inappropriateness of the metrics of accuracy, precision, recall, and JAC are analyzed. A simulation of experiment that an unbiased Gaussian estimator achieving the Fisher information (therefore achieving CRLB) localizes emitter locations frame by frame independently is carried out. The simulation result demonstrates the superior sensitivity of RMSMD to a quality change over the other four metrics. The RMSMD and visual quality of the information-achieving localization nanoscopy image can be used as a benchmark for all localization nanoscopy images and algorithms. As a universal and objective metric, RMSMD can be broadly employed in applications to measure the fitness between two sets of points.

## Method

**Definitions.** Consider two sets of $n$-dimensional points $S = \{s_i, i = 1, \ldots, M\}$ and $X = \{x_i, i = 1, \ldots, N\}$ where $s_i$'s or $x_i$'s are unnecessary to be distinct. The mean square minimum distance (MSMD) between $X$ and $S$ is defined as

$$D^2(X, S) = \frac{1}{|X| + |S|} \left( \sum_{s \in S} \min_{x \in X} \|x - s\|^2 + \sum_{x \in X} \min_{s \in S} \|s - x\|^2 \right) \tag{1}$$

where $|\cdot|$ is the number of elements in a set and $\|\cdot\|$ is the $l_2$ norm or the Euclidean distance between two points. Then their root mean square minimum distance (RMSMD) is $D(X, S)$. If a point in one set has the same minimum distance to multiple points in the other set, one of the multiple points is arbitrarily chosen and the RMSMD does not depend on the choice.

In the definition, $\min_{x \in X} \|x - s\|^2$ for $s \in S$ is the minimum square distance from the point $s$ to the set $X$, and $\min_{s \in S} \|s - x\|^2$ for $x \in X$ is the minimum square distance from the point $x$ to the set $S$. Therefore, $D^2(X, S)$ is the average minimum square distance from a point in $X$ to $S$ and from a point in $S$ to $X$. Because of this, $D(X, S)$ evaluates how well the two sets $X$ and $S$ averagely, locally, and mutually fit to each other. The smaller the RMSMD, the better the fitness between the two sets. Since $D(X, S)$ depends only on $X$ and $S$ regardless of how they are produced, RMSMD is a universal and objective metric and can be extensively applied in various applications.

It is worth noting that both summation terms in the definition of RMSMD are necessary; otherwise it may result in an awkward measurement of fitness between $X$ and $S$. For example, suppose that only the second summation term in Eq. (1) were considered in the definition of RMSMD. When $M \geq 2$, $N = 1$, and $x_1 = s_1 \neq s_2$ the second term is equal to zero, which would imply a perfect fitness. However, $X$ and $S$ could be actually significantly different in the case that $s_2$ is far away from $s_1$. Such a difference is taken into account by the first summation term in Eq. (1).

The Voronoi cell of $s_i \in S$ is defined by $V(s_i) = \{x \in \mathbb{R}^n, \|x - s_i\| \leq \|x - s_j\|, j \neq i\}$ that contains all the points in $\mathbb{R}^n$, whose distance to $s_i$ is not greater than to any other $s_j$ for $j \neq i$. Any two adjacent Voronoi cells $V(s_i)$ and $V(s_j)$ are separated by the hyperplane that contains the middle point of and is orthogonal to the line segment connecting $s_i$ and $s_j$. The $n$-dimensional space $\mathbb{R}^n$ is partitioned by the $M$ Voronoi cells $V(s_i)$. Similarly, the Voronoi cell $V(x_i)$ for $x_i \in X$ is defined. The $n$-dimensional space $\mathbb{R}^n$ is also partitioned by the $N$ Voronoi cells $V(x_i)$. In terms of the Voronoi cells, the MSMD can be written as

$$D^2(X, S) = \frac{1}{|X| + |S|} \left( \sum_{x \in X} \sum_{s \in S \cap V(x)} \|x - s\|^2 + \sum_{s \in S} \sum_{x \in X \cap V(s)} \|s - x\|^2 \right) \tag{2}$$

We define a *kernel* point and a kernel set as follows. $x^*$ is the point in $X$ that has the minimum distance to a point $s^*$ in $S$, i.e., $x^* = \arg \min_{x \in X} \|x - s^*\|$; meanwhile, the point $s^*$ is also the point in $S$ that has the minimum distance to $x^*$, i.e., $s^* = \arg \min_{s \in S} \|s - x^*\|$. Then $x^*$ is said to be a kernel point of $X$ with respect to $S$. It is clear that $s^*$ is also a kernel point of $S$ with respect to $X$. $(x^*, s^*)$ is said to be a pair of kernel points in $X \times S$. The set of all and only kernel points of $X$ is called the kernel set of $X$ with respect to $S$. Similarly, the kernel set of $S$ with respect to $X$ contains only the kernel points of $S$.

Figure 1 shows an example of 2D point sets $S$ and $X$, the points reaching the minimum distances, the kernel points, the kernel sets, and the Voronoi cells.

**Application to localization nanoscopy.** RMSMD can be applied to localization nanoscopy imaging in various situations. $S$ can be a set of emitter locations and $X$ is a set of locations estimated from a sequence of data frames generated by the emitters. For 2D imaging of static samples, $s_i$'s and $x_i$'s are 2D in the lateral plane. For 3D imaging of static samples, $s_i$'s and $x_i$'s are 3D, and their first two coordinates are lateral and their third coordinate is axial. For 2D imaging of a dynamic process, $s_i$'s and $x_i$'s are 3D, and their first two coordinates are lateral and their third coordinate is the time of frame. For 3D imaging of a dynamic process, $s_i$'s and $x_i$'s are 4D; their first two coordinates are lateral, their third coordinate is axial, and their fourth coordinate is the time of frame.

 

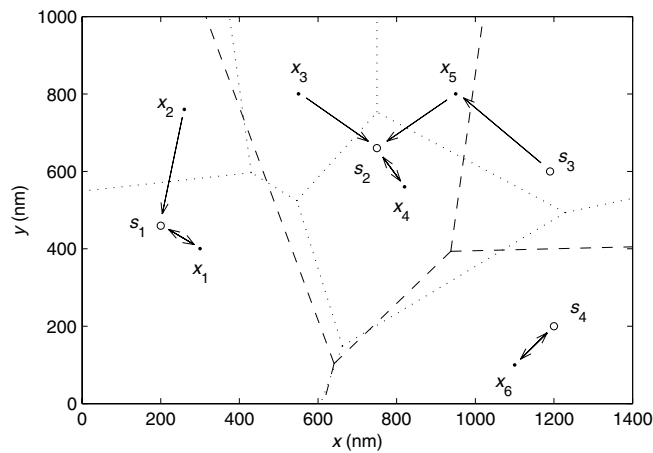

**Figure 1.** Two sets of 2D points $S = \{s_1, s_2, s_3, s_4\} = \{(200, 460), (750, 660), (1190, 600), (1200, 200)\}$ denoted by circles and $X = \{x_1, x_2, x_3, x_4, x_5, x_6\} = \{(300, 400), (260, 760), (550, 800), (820, 560), (950, 800), (1100, 100)\}$ denoted by dots. A line starts from a point of one set and directs by an arrow to the minimum-distance point of the other set. A pair of kernel points of $S$ and $X$ are directed by a line, of which both ends are arrows. The kernel sets of $X$ and $S$ are $X^* = \{x_1, x_4, x_6\}$ and $S^* = \{s_1, s_2, s_4\}$, respectively, and the pairs of kernel points are $(x_1, s_1), (x_4, s_2),$ and $(x_6, s_4)$. The Voronoi cells $V(s_i)$ and $V(x_j)$ are bounded by the dashed and dotted lines, respectively. The points of $X$ in the Voronoi cells of $s_i \in S$ are $X \cap V(s_1) = \{x_1, x_2\}, X \cap V(s_2) = \{x_3, x_4, x_5\}, X \cap V(s_3) = \varnothing$, and $X \cap V(s_4) = \{x_6\}$. The points of $S$ in the Voronoi cells of $x_i \in X$ are $S \cap V(x_1) = \{s_1\}, S \cap V(x_2) = \varnothing$, $S \cap V(x_3) = \varnothing, S \cap V(x_4) = \{s_2\}, S \cap V(x_5) = \{s_3\}$, and $S \cap V(x_6) = \{s_4\}$.

Given a set of emitter locations, RMSMD can evaluate the quality of a nanoscopy image estimated by a localization algorithm, and therefore evaluate the performance of the algorithm. The evaluation can be based on all the locations estimated from the entire sequence of acquired data frames. There are two categories of localization algorithms in the literature. In the first category, a localization algorithm like the Bayesian localization algorithm[10] jointly uses the entire sequence of data frames to estimate one location for each emitter, and therefore $M$ and $N$ are in the same magnitude. In contrast, belonging to the second category, most localization algorithms[1–3] localize the activated emitters frame by frame independently and do not identify an estimated location with a particular emitter. Since each emitter is activated a number of times in the sequence of data frames, these algorithms produce multiple estimated locations for each emitter. The nanoscopy image consists of the estimated locations from all data frames and therefore the number of estimated locations is much larger than the number of emitters, i.e., $N \gg M$. For example, if each emitter is activated $K$ times on average, then $N \cong KM$ on average where $K$ can be as large as 50 for some kinds of emitters[11]. For the second category of localization algorithms, RMSMD can also be applied to evaluate the quality of emitter locations estimated from each single frame. In this case, $S$ and $X$ are the set of locations of the activated emitters in a frame and the set of their estimates, respectively, and then $N$ and $M$ differ slightly.

A sample drifting may cause additional translation and rotation for the entire set of estimated locations $X$ and then the RMSMD $D(X, S)$ might be unnecessarily increased. However, such an increase can be eliminated by performing the converse translation and rotation for the entire set of estimated locations. Because of this, throughout we assume that such additional translation and rotation have been removed from $X$.

## RMSMD Properties

A number of properties of RMSMD are presented in this section. Some of them are common to a distance metric just as RMSMD, and some others are unique to RMSMD.

### Properties common to a distance metric.

We present the following properties of RMSMD that are common to a distance metric. The proof of them is straightforward and is omitted.

*Property 1 (Non-negativity and identity)*: $D(X, S) \geq 0$ where the equality holds if and only if $X$ and $S$ differ only by a permutation of indices of their points.

*Property 2 (Symmetry)*: $D(X, S) = D(S, X)$.

*Property 3 (Triangle inequality)*: $D(X, S) \leq D(X, Y) + D(Y, S)$.

*Property 4 (Invariance to translation and rotation)*: $D(X, S) = D(X', S')$ where $X'$ and $S'$ are obtained by the same translation and rotation from $X$ and $S$, respectively.

*Property 5 (Continuity)*: $D(X, S)$ is a continuous function of all $x \in X$ and $s \in S$.

The continuity of RMSMD is an important property when applied to localization nanoscopy images.

### Properties of kernel points.

In addition to the properties common to a distance metric, RMSMD also has the following unique properties.

*Property 6 (A pair of kernel points)*: $(x^*, s^*)$ is a pair of kernel points if and only if they are located in each other's Voronoi cell, that is, $x^* \in V(s^*)$ and $s^* \in V(x^*)$.

*Property 7 (Removal of far non-kernel points)*: Suppose that $X' \subset X$ is a subset of non-kernel points such that (i) $S \cap V(x) = \varnothing$ for any $x \in X'$, (ii) the average of square minimum distances $\min_{s \in S} \|s - x\|^2$ for all $x \in X'$ is larger than the average of the remaining square distances in $D(X, S)$. Then $D(X - X', S) < D(X, S)$.

If $X^*$ and $S^*$ are kernel sets for each other, then

$$D^2(X^*, S^*) = \frac{1}{|S^*|} \sum_{i=1}^{|S^*|} \|x_i^* - s_i^*\|^2$$

(3)

where $(x_i^*, s_i^*)$ is the $i$th pair of kernel points.

*Property 8 (Kernel points)*: Let $X^* \subseteq X$ and $S^* \subseteq S$ be the sets of kernel points of $X$ and $S$, respectively. If the average of square distances between each pair of kernel points $(x^*, s^*) \in X^* \times S^*$ is smaller than the average of the remaining square distances in $D(X, S)$, then $D(X^*, S^*) < D(X, S)$.

Property 7 indicates that removing a subset of non-kernel points from $X$ that are far away from the kernel points of $S$ can decrease RMSMD. Property 8 further indicates that all non-kernel points can be removed to reduce RMSMD if the average of square distances between each pair of the kernel points is smaller than the average of the remaining square distances. Property 7 and Property 8 are proved in the Appendix.

For example, in Fig. 1 the MSMD between $S$ and $X$ is

$$\begin{aligned}
D^2(X, S) &= \frac{1}{10}(2\|x_1 - s_1\|^2 + 2\|x_6 - s_4\|^2 + 2\|x_4 - s_2\|^2 \\
&\quad + \|x_3 - s_2\|^2 + \|x_5 - s_2\|^2 + \|x_5 - s_3\|^2 + \|x_2 - s_1\|^2) \\
&= 40740 \ (\text{nm}^2).
\end{aligned}$$

(4)

For $X' = \{x_2, x_3\}$, the two conditions in Property 7 are satisfied in that (i) $S \cap V(x_2) = \varnothing, S \cap V(x_3) = \varnothing$ and (ii) $(\|x_3 - s_2\|^2 + \|x_2 - s_1\|^2)/2 = 76600 \ (\text{nm}^2)$ which is larger than

$$\begin{aligned}
D^2(X - X', S) &= \frac{1}{8}(2\|x_1 - s_1\|^2 + 2\|x_6 - s_4\|^2 \\
&\quad + 2\|x_4 - s_2\|^2 + \|x_5 - s_2\|^2 + \|x_5 - s_3\|^2) \\
&= 31775 \ (\text{nm}^2).
\end{aligned}$$

(5)

It confirms Property 7 that removing $X'$ from $X$, we obtain

$$D^2(X - X', S) < D^2(X, S).$$

(6)

The kernel sets are $X^* = \{x_1, x_4, x_6\}$ and $S^* = \{s_1, s_2, s_4\}$. It confirms Property 8 that

$$\begin{aligned}
D^2(X, S) > D^2(X^*, S^*) &= \frac{1}{3}(\|x_1 - s_1\|^2 + \|x_6 - s_4\|^2 + \|x_4 - s_2\|^2) \\
&= 16166 \ (\text{nm}^2).
\end{aligned}$$

(7)

Looking at Fig. 1, it is clear that $X - X'$ is a better fitness of $S$ than $X$, and $X^*$ and $S^*$ fit better to each other than $X$ and $S$ do, which are predicted by the corresponding RMSMDs. In return, it implies that RMSMD is a rational measurement of mutual fitness between two sets of points.

In localization nanoscopy, a majority of localization algorithms in the literature[1,3] localize emitters independently frame by frame and produce a number of estimated locations for each emitter. Most of the estimated locations are far away from the ground-truth locations and therefore can be removed to reduce RMSMD and improve the quality of nanoscopy image according to Property 7. As implied by Property 8, keeping only the kernel estimated locations can minimize RMSMD. Hence, an algorithm that without knowing the ground-truth locations can identify a number of non-kernel estimated locations with certain probability is desirable.

## Accuracy, precision, recall, and JAC

Accuracy, precision, recall, and JAC are currently employed in the literature as quality metrics for localization nanoscopy images[1,3,8,9]. To demonstrate the advantages of RMSMD over these four metrics, in this section we investigate their ambiguity, discontinuity, and failure in distinguishing qualities of different images in certain conditions.

Accuracy, precision, recall, and JAC all are defined on the basis of the true-positive regions of emitter locations $s_i$'s. The area within a circle centered at $s_i$ is considered as the true-positive region of $s_i$ and the rest is considered as its false-positive region. The radius of the circle is equal to the FWHM of PSF in an optical system. An estimated location $x_i$ is categorized as a true positive (TP) point if it is located in any true-positive region of $s_i$. $x_i$ is categorized as a false positive (FP) point if it is not located in any true-positive region of $s_i$. $s_i$ is categorized as a false negative (FN) point if its true-positive region contains no $x_i$. Based on the sets of TP, FP, and FN points, accuracy, precision, recall, and JAC are defined as[1,3]

$$\text{Accuracy} = \sqrt{\frac{1}{|\text{TP}|} \sum_{i=1}^{|\text{TP}|} a_i^2},$$

(8)

$$\text{Precision} = \frac{|\text{TP}|}{|\text{TP}| + |\text{FP}|}, \tag{9}$$

$$\text{Recall} = \frac{|\text{TP}|}{|\text{TP}| + |\text{FN}|}, \tag{10}$$

$$\text{JAC} = \frac{|\text{TP}|}{|\text{TP}| + |\text{FP}| + |\text{FN}|}, \tag{11}$$

respectively, where $a_i$ for $i = 1, \ldots, |\text{TP}|$ is the distance from a true-positive point $x_k$ to a nearest $s_j$.

According to their definitions, accuracy, precision, recall and JAC have the following substantial drawbacks.

(i) *FP ambiguity*: The locations of FP points are absent in these four metrics. If the FP points move and keep in the FP region, then all these four metrics remain unchanged and falsely imply an unchanged quality of $X$.

(ii) *FN ambiguity*: The locations of FN points are absent in these four metrics. If the FN points move and still contain no $x_i$ in their TP regions, then all these metrics remain unchanged and falsely imply an unchanged quality of $X$.

(iii) *Discontinuity in X*: If an $x_i$ moves and crosses the boundary of a TP region, which switches $x_i$ from a TP point to a FP point or conversely, then all these four metrics discontinuously change.

(iv) *Discontinuity in S*: If an $s_j$ moves such that an $x_j$ crosses the boundary of $s_j$'s TP region, then all these four metrics discontinuously change.

(v) *Subjectiveness and non-universality*: The TP region depends on the PSF FWHM of an optical system in an experiment and so do all the four metrics. Hence, the four metrics are subjective and non-universal.

The FP and FN ambiguities imply that these metrics fail to distinguish the qualities of different nanoscopy images in certain conditions.

*Remarks* (Estimation vs. detection): In a statistical decision of the value of a variable through observations, there are two basic problems: one is estimation and the other is detection[12]. In an estimation problem, the variable with an unknown value is defined on a continuous support and so is continuous and can take uncountably infinitely many possible values. A device to make a decision on the unknown value in an estimation problem is called an estimator. On the other hand, in a detection problem, the variable with an unknown value can only take a finite number of possible values. A device to make a decision on the unknown value in a detection problem is called a detector. In a binary detection problem a detector can only choose one of two possible values, say true or false. In localization nanoscopy, since the emitter locations are defined on $\mathbb{R}^n$, i.e., a real space with dimension $n = 2$ or 3, the localization of an emitter is an estimation problem but not a detection problem. For the same reason, how well $X$ (a set of estimated emitter locations) and $S$ (the set of ground-truth emitter locations) mutually fit to each other is a problem to measure the performance of an estimator instead of a detector.

It is worthy to clarify what accuracy, precision, recall, and JAC are appropriate and inappropriate to measure. Precision and recall are suitable for measuring the performance of a detector in a binary detection problem. When applied to measuring the performance of emitter localization – an estimation problem, $x_i$'s have to be divided into two groups, TP and FP. To this end, in refs[1,3] the circle centered at $s_i$ with the radius of PSF FWHM in an optical system is used to divide the space into the true-positive and false-positive regions. Such a division depends on the optical system in an experiment, and so is somewhat subjective and arbitrary, and is not universal as well. JAC is originally suitable for measuring the fitness of two continuous sets consisting of uncountably infinitely many points. However, $X$ and $S$ in localization nanoscopy are two discrete sets each consisting of countably many locations, and so their mutual fitness is inappropriate to be measured by JAC. The original accuracy is suitable to measure the global error between an estimated value and its ground-truth value. When accuracy is applied to localization nanoscopy, the partition set $X_i$ of $X$ for each ground-truth location $s_i$ must be known. In practice, however, only the entire set $X$ of estimated emitter locations is known but its partition sets $X_i$'s are unknown. To circumvent this problem, in refs[1,3] those $x_j$'s located in the true-positive region of $s_i$ are considered as estimates of $s_i$ and are taken into account in calculation of accuracy in Eq. (8); and the other estimated locations that are not located in the true-positive region of any $s_i$ are ignored. Because of this, the accuracy based on the true-positive region cannot properly measure the overall fitness between $X$ and $S$. Even when the partition $X_i$'s were known, the accuracy in Eq. (8) would only measure the error between a ground-truth location and its estimates and cannot properly measure the average, local, and mutual fitness between $X$ and $S$ that truly needs to be measured.

In summary, precision and recall are not suitable for performance evaluation of an estimator. The modified JAC and the modified accuracy still fail to properly measure the mutual fitness between two discrete sets of points. In contrast, as a universal and objective metric, RMSMD can properly measure the average, local, and mutual fitness between two sets of points.

## Numerical Examples

In this section, the performances of RMSMD and the four metrics of accuracy, precision, recall, and JAC are compared by two numerical examples.

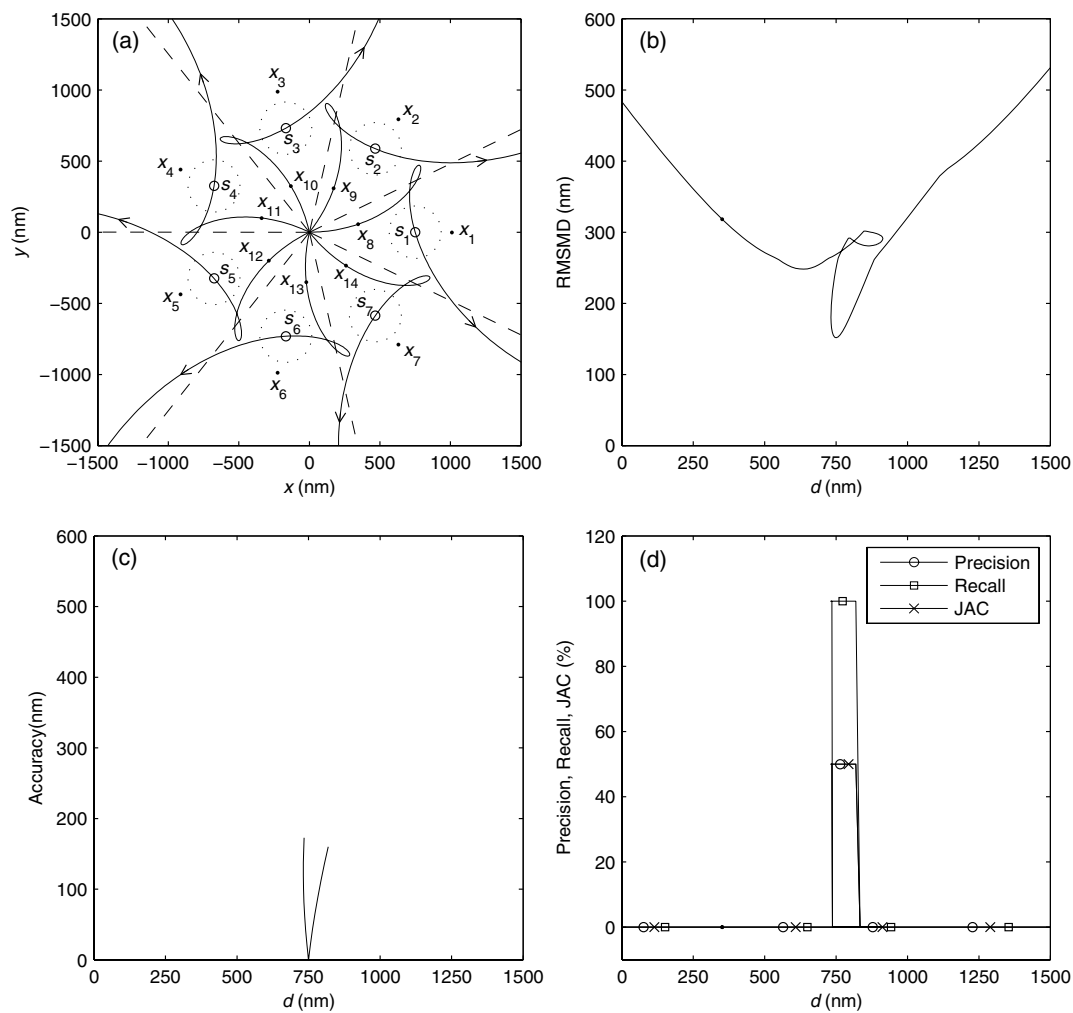

**Figure 2.** Performance comparison of RMSMD with the other four metrics. (**a**) Two sets of 2D points, $S$ and $X$. (**b**) RMSMD, (**c**) accuracy, (**d**) precision, recall, and JAC as functions of $d$. The dots in (**b**), (**d**) correspond to the dotted locations of $x_8, \ldots, x_{14}$ in (**a**). $d$ is the distance from the origin to anyone of $x_8, \ldots, x_{14}$.

**A variable set of points.** Consider Fig. 2(a). $S$ consists of seven 2D emitter locations $s_1, \ldots, s_7$ that are denoted by the little circles and equally spaced on a circle centered at the origin with a radius of 750 nm. The Voronoi cells of $s_i$'s are bounded by the dashed lines. As an estimate of $S$, $X$ consists of 14 locations $x_1, \ldots, x_{14}$ denoted by the dots. $x_i$ for $i = 1, \ldots, 7$ are fixed and equally spaced on a circle centered at the origin with the radius of 1012 nm, and the origin, $s_i$, and $x_i$ are on the same line. We intend to investigate how the metrics change with a continuously changing $X$. To this end, $x_8, \ldots, x_{14}$ move along their equally-spaced trajectories denoted by the solid lines in the direction denoted by the arrows. $x_8, \ldots, x_{14}$ move at the same speed so that at any moment they are also equally spaced on a circle centered at the origin. Their locations at a particular instant are denoted by the dots. To calculate accuracy, precision, recall, and JAC, FWHM = 184 nm is taken. Inside the doted circle centered at $s_i$ with the radius of FWHM is the true-positive region of $s_i$. Outside these circles, $x_1, \ldots, x_7$ are in the false-positive region. Since $X$ is a function of the distance $d$ between the origin and anyone of $x_8, \ldots, x_{14}$, all metrics are also functions of $d$. The metrics versus $d$ corresponding to the moving $x_8, \ldots, x_{14}$ are shown in Fig. 2(b)–(d).

As shown in Fig. 2(b), RMSMD continuously properly measures the fitness between $X$ and $S$ in all cases as $X$ continuously changes.

In contrast, as shown in Fig. 2(c) and (d), accuracy, precision, recall and JAC present drawbacks. First, they are discontinuous with respect to $d$. Second, when $x_8, \ldots, x_{14}$ are in the false-positive region, accuracy is undefined, and precision, recall, and JAC are constantly equal to zero. That is, they all fail to distinguish the qualities of continuously changing $X$ as $x_8, \ldots, x_{14}$ move along the corresponding segment of trajectories. Third, when $x_8, \ldots, x_{14}$ are in the true-positive region, precision, recall, and JAC are always equal to constants 50%, 100%, and 50%, respectively, and fail to distinguish the qualities of continuously changing $X$. Finally, in the particular case when $x_8, \ldots, x_{14}$ are coincided with $s_1, \ldots, s_7$, respectively, accuracy is equal to zero, which implies an illusive perfect fitness between $X$ and $S$ but actually $x_1, \ldots, x_7$ are not taken into account. These drawbacks are inherently incurred by the FP and FN ambiguities, FP and FN discontinuities, and inappropriateness of accuracy, precision, recall and JAC as quality metrics.

 6

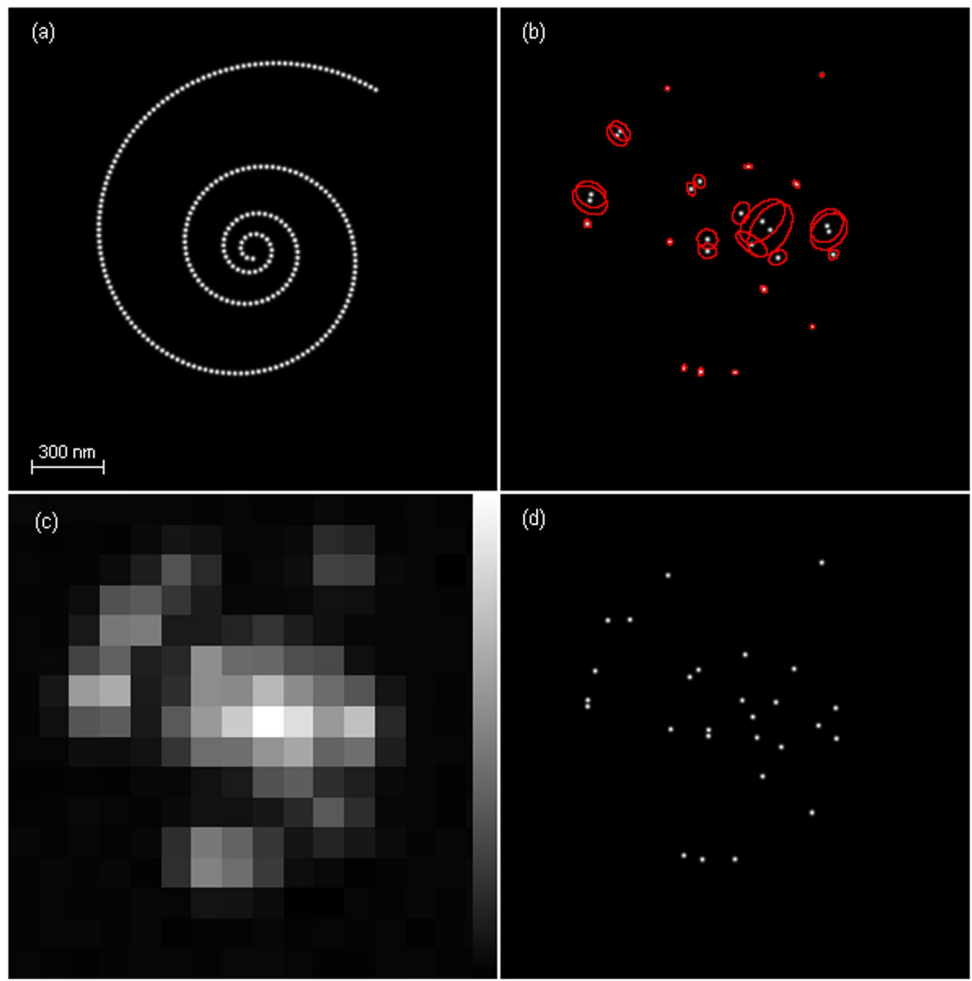

**Figure 3.** Emitter locations and their estimates. (**a**) 250 emitters located on a 2D helix with the equal spacing of 25 nm. (**b**) 27 emitters that are activated in a data frame when $r_{00} = 0.92$. The red ellipses indicate the 95%-probability regions of the activated emitters. (**c**) A data frame with the 27 activated emitters in (**b**). (**d**) Corresponding to the activated emitters in (**b**), 27 estimated emitter locations by an unbiased Gaussian information-achieving estimator.

**Nanoscopy imaging with an unbiased Gaussian information-achieving estimator.** We simulate an experiment of 2D localization nanoscopy imaging where an unbiased Gaussian estimator that achieves the Fisher information and CRLB localizes activated emitters frame by frame independently. The RMSMD of an information-achieving nanoscopy image can benchmark localization nanoscopy images obtained by other algorithms. The data frame model, Fisher information matrix, and CRLB are defined and presented in detail in ref.[7] to which the reader is referred.

*Data frame model.* As shown in Fig. 3(a), $M = 250$ emitters are located on a 2D helix with the adjacent emitter distance equal to 25 nm. Each data frame is generated according to the model in ref.[7]. The optical system has a numerical aperture of $n_a = 1.4$ and the fluorescence wavelength of emitters is $\lambda = 540$ nm. Then the standard deviation of an Airy PSF can be approximately calculated by Eq. (49) in ref.[7] and is equal to $\sigma = 78.26$ nm, and accordingly FWHM $= 2\sqrt{2ln2}\,\sigma = 184.28$ nm. A Gaussian PSF with zero mean is used to approximate the Airy PSF with the same standard deviation $\sigma$. An activated emitter in a frame emits a number of photons that is Poisson distributed with the mean of $I$ photons per second. Each photon is detected by the camera with a probability of $\eta$. The mean number of photons that are emitted from an activated emitter and are successfully detected by a camera is equal to $I\eta = 300000$ photons per second. The frame rate is $1/\Delta t = 100$ frames per second. Then the mean number of detected photons is equal to $I\eta\Delta t = 3000$ per frame per emitter. The additive Poisson noise and Gaussian noise are presented in the data frames[7]. The signal to Poisson noise ratio (SPNR) and the signal to Gaussian noise ratio (SGNR) are SPNR $= 0.2$ and SGNR $= 0.3$ μm² per emitter, respectively. The frame size is $L_x \times L_y = 2048 \times 2048$ nm² and the pixel size is $\Delta_x \times \Delta_y = 128 \times 128$ nm², and then each frame has $K_x \times K_y = 16 \times 16$ pixels. The photon emissions of all emitters, Poisson noise, and Gaussian noise are all independent.

*Emitter activation process.* The activation process of an emitter can be modeled by a Markov chain[10,13]. We consider a more general chain than those of refs[10,13]. Specifically, each emitter in a data frame is independently activated and the state of activation follows a Markov chain taking states 0, 1, 2, 3, 4. State 0 means no activation and the other four states mean activation. The transition probability from state $i$ to state $j$ is denoted by $r_{ji}$. In the simulation, $r_{00} > 0$ is variable, and $r_{01} = 0.5$, $r_{02} = 0.7$, $r_{03} = 0.8$, and $r_{04} = 1$ are fixed; correspondingly, $r_{10} = 1 - r_{00}$, $r_{21} = 1 - r_{01}$, $r_{32} = 1 - r_{02}$, and $r_{43} = 1 - r_{03}$, and all other transition probabilities are zero. It is clear that the Markov chain is irreducible, aperiodic, and positive recurrent and therefore has a unique stationary probability distribution[14]. The stationary probability of state 0 is equal to $p_0 = 1/\left(1 + \sum_{j=1}^{4} \prod_{k=0}^{j-1} r_{k+1,k}\right)$ and then the stationary probability of activation is equal to $p_a = 1 - p_0$. $N$ data frames are acquired. The number of activated emitters per frame has a Binomial distribution with mean $M_a = p_a M$, the number of activations per emitter is Binomial distributed with mean $p_a N$, and the total number of activations of all emitters in $N$ data frames is Binomial distributed with mean $p_a MN$. When $r_{00}$ is given, these mean numbers can be calculated. In the simulation, the mean number of activations per emitter in $N$ frames is fixed to be $p_a N = 30$. In the simulation, $r_{00} = 0.98, 0.97, 0.96, 0.95, 0.94$, 0.93, 0.92 are considered. Correspondingly, the emitter activation probability is $p_a = 0.0325, 0.0480, 0.0630$, 0.0775, 0.0916, 0.1052, 0.1185, the mean number of activated emitters per frame is $M_a = 8.1, 12.0, 15.7, 19.4, 22.9$, 26.3, 29.6, the corresponding average density of activated emitters in a frame is equal to 1.94, 2.86, 3.75, 4.62, 5.46, 6.27, 7.06 emitters per $\mu m^2$, and the total number of frames is $N = 922, 625, 476, 387, 327, 285, 253$, respectively. In all cases, the mean number of activations of all emitters in $N$ data frames is equal to $p_a MN = 30 \times 250 = 7500$. To cover a broad range of emitter density, the average density of activated emitters in a data frame is purposely set up higher than that in practice[1,3].

*Unbiased Gaussian estimator achieving the Fisher information.* In each data frame, a random number $K$ of emitters are activated as shown in Fig. 3(b). Denote by $\theta_i = (x_i, y_i)^T$ the location coordinates of the $i$th activated emitter and by $\theta = (\theta_1^T, \ldots, \theta_K^T)^T$ the location coordinates of all activated emitters. Given the activated emitters in a data frame and the data frame model, the Fisher information matrix $\mathbf{F}$ of the location coordinates of the activated emitters $\theta$ can be calculated[7]. The Fisher information determined by $\mathbf{F}$ measures the information about $\theta$ provided by the data frame on average. The diagonal elements of $\mathbf{F}^{-1}$ denoted by $\sigma_{x1}^2, \sigma_{y1}^2, \cdots, \sigma_{xK}^2, \sigma_{yK}^2$ are called the CRLB[6,7] that are the smallest variances possibly achievable by all unbiased estimators for $x_1, y_1, \ldots, x_K, y_K$, respectively. An unbiased estimator that achieves the Fisher information, i.e., whose Fisher information matrix is $\mathbf{F}$, also achieves the CRLB.

To generate an unbiased Gaussian estimator $\hat{\theta}$ achieving the Fisher information, we consider a Gaussian estimator $\hat{\theta}$ with mean $\theta$ and covariance matrix $\mathbf{F}^{-1}$, i.e. $\hat{\theta} \sim N(\theta, \mathbf{F}^{-1})$. It is straightforward to verify that the Fisher information matrix of $\hat{\theta}$ is equal to $\mathbf{F}$. Therefore, $\hat{\theta}$ is unbiased and Gaussian and achieves the Fisher information and the CRLB as well. Denote by $\mathbf{F}^{-1} = \mathbf{U}\mathbf{\Lambda}\mathbf{U}^T$ the eigendecomposition of $\mathbf{F}^{-1}$ where $\mathbf{\Lambda}$ is the diagonal matrix of eigenvalues and $\mathbf{U}$ is the matrix of corresponding eigenvectors of $\mathbf{F}^{-1}$. Let $g$ be a $2K$-dimensional standard Gaussian random vector with zero mean and unit covariance matrix. Then an unbiased Gaussian estimator of $\theta$ that achieves the Fisher information and CRLB is obtained by

$$\hat{\theta} = \theta + \mathbf{U}\mathbf{\Lambda}^{0.5}g. \tag{12}$$

It is clear that $\hat{\theta} \sim N(\theta, \mathbf{F}^{-1})$. Accordingly, the $i$th 2D vector of $\hat{\theta} = \left(\hat{\theta}_1^T, \ldots, \hat{\theta}_K^T\right)^T$, i.e., $\hat{\theta}_i = (\hat{x}_i, \hat{y}_i)^T$, is an unbiased Gaussian estimator of $\theta_i = (x_i, y_i)^T$ with mean $\theta_i$ and covariance matrix $\mathbf{F}_i^{-1}$ where $\mathbf{F}_i^{-1}$ is the $i$th $2 \times 2$ diagonal block matrix of $\mathbf{F}^{-1}$. That is, $\hat{\theta}_i \sim N(\theta_i, \mathbf{F}_i^{-1})$. The diagonal elements of $\mathbf{F}_i^{-1}$, i.e., $\sigma_{xi}^2$ and $\sigma_{yi}^2$, are the CRLBs of $x_i$ and $y_i$, respectively. Therefore, $\hat{\theta}_i$ is an unbiased Gaussian estimator achieving the CRLB of $\theta_i$. A realization of $\hat{\theta}$ can be obtained with given a realization of $g$, say a pseudorandom vector generated by the embedded function *randn* in MATLAB. The realization of $\hat{\theta}_i$ is one of the estimated locations from a data frame and is shown in the final nanoscopy image as presented in Fig. 4.

*95%-probability region.* It is interesting to know where $\hat{\theta}_i$ is mostly located. To this end, we define the 95%-probability region of $\hat{\theta}_i$ as the region where $\hat{\theta}_i$ is mostly located and the total probability is 0.95. Since $\hat{\theta}_i$ is Gaussian distributed with mean $\theta_i$ and covariance matrix $\mathbf{F}_i^{-1}$, it is clear that the 95%-probability region is given by the ellipsoid

$$(\hat{\theta}_i - \theta)^T \mathbf{F}_i (\hat{\theta}_i - \theta_i) \leq R^2 \tag{13}$$

where $R > 0$ is a constant determined by $\Pr[(\hat{\theta}_i - \theta_i)^T \mathbf{F}_i(\hat{\theta}_i - \theta_i) \leq R^2] = 0.95$. Denote by $\mathbf{F}_i^{-1} = \mathbf{U}_i \mathbf{\Lambda}_i \mathbf{U}_i^T$ the eigendecomposition of $\mathbf{F}_i^{-1}$ where $\mathbf{\Lambda}_i$ is the diagonal matrix of eigenvalues and $\mathbf{U}_i$ is the matrix of corresponding eigenvectors of $\mathbf{F}_i^{-1}$. Define $v_i = \mathbf{\Lambda}_i^{-0.5}\mathbf{U}_i^T(\hat{\theta}_i - \theta_i)$ and then $R$ satisfies $\Pr(\|v_i\| \leq R) = 0.95$. Since $v_i \sim N(0, \mathbf{I})$ is Gaussian distributed with zero mean and unit covariance matrix, $\|v_i\|$ is Chi distributed and its number of degrees of freedom is equal to $n$ where $n$ is the dimension of $v_i$. Therefore, $P(n/2, R^2/2) = 0.95$ where $P$ is the regularized gamma function, the cumulative distribution function of $\|v_i\|$. When $n = 2$, $P$ becomes the Rayleigh distribution[15], i.e., $1 - \exp(-R^2/2) = 0.95$ and hence,

$$R = \sqrt{-2 \ln 0.05} \tag{14}$$

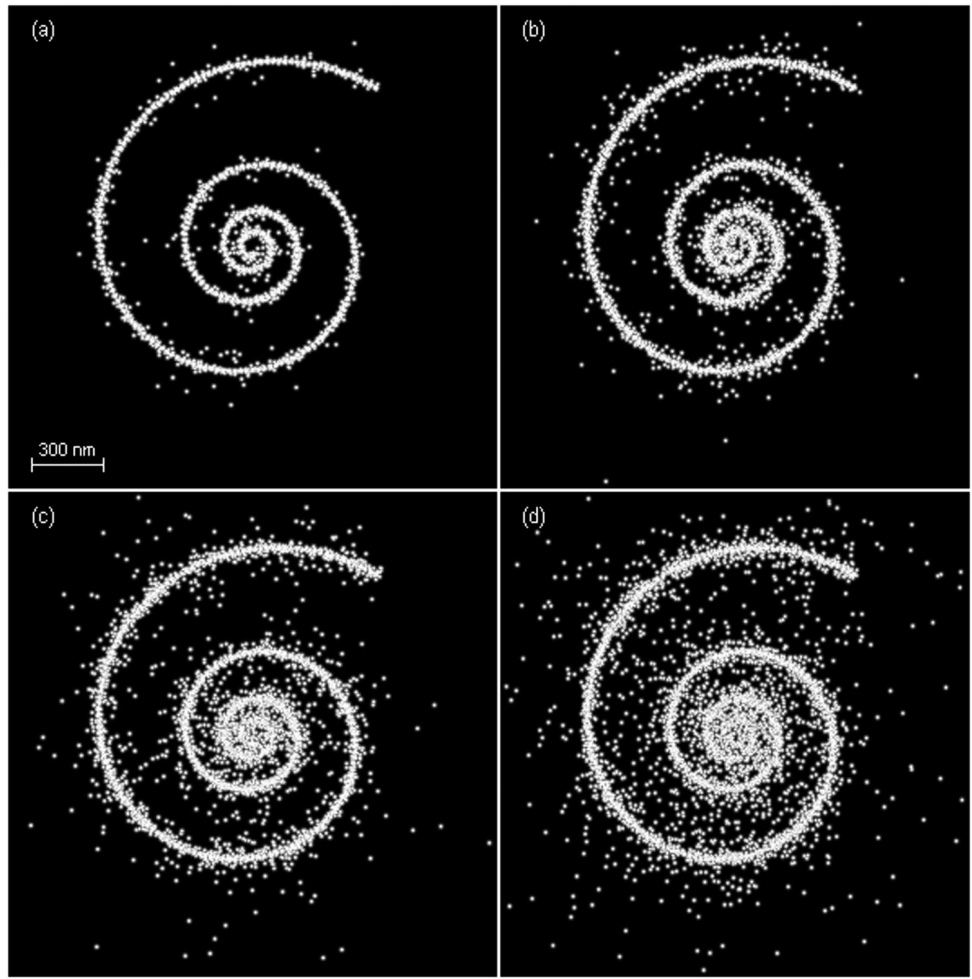

**Figure 4.** Unbiased Gaussian information-achieving nanoscopy images versus $M_a$, the mean number of activated emitters per frame. (**a**) $M_a = 8.1$. (**b**) $M_a = 15.7$. (**c**) $M_a = 22.9$. (**d**) $M_a = 29.6$. Some estimated locations are located outside the region of image and therefore the true visual qualities of (**c**) and (**d**) are worse than those shown here.

or $R \cong 2.448$. The ellipse of the 95%-probability region is given by $\theta_i + R U_i \Lambda_i^{0.5} q_i$, where $q_i$ is on the unit circle of $\|q_i\| = 1$, and can be drawn in a two-dimensional image. The larger the 95%-probability region, the larger the estimation error of $\hat{\theta}_i$.

For the activated emitters in Fig. 3(b), their 95%-probability regions of the information-achieving estimators are shown. As can be seen, the emitters located close to each other with their PSFs severely overlapped are suffered from the significant inter-emitter interference as discussed in ref.[7] and have a large 95%-probability region, and their estimation errors in $x$ and $y$ directions may be considerably correlated due to a large difference between the two eigenvalues. A data frame generated according to the data model is shown in Fig. 3(c). By means of Eq. (12) the estimated emitter locations by the unbiased estimator achieving the Fisher information are shown in Fig. 3(d). The nanoscopy image consists of all the emitter locations estimated from the $N$ data frames as shown in Fig. 4(a)–(d). Since all activated emitters are localized frame by frame independently, the nanoscopy images achieve the Fisher information and CRLB in the sense that the estimated emitter locations achieve the Fisher information and CRLB in each corresponding data frame.

Indexed by the mean number of activated emitters per frame $M_a$, the RMSMD, accuracy, precision, recall, and JAC of the nanoscopy images are shown in Fig. 5(a) and (b). As the mean number of activated emitters per frame increases, the RMSMD increases exponentially fast, which confirms the result that the CRLB exponentially increases as the emitter density increases[7], indicating that the quality of the nanoscopy image decreases fast. The fast increase of both RMSMD and CRLB is due to the estimated locations spread away from the ground-truth locations caused by the increasing severity of the inter-emitter interference. Subjectively, the visual quality of the nanoscopy image indeed deteriorates fast accordingly. In contrast, the accuracy, precision, recall, and JAC all change slightly and are insensitive to the quality change. The result demonstrates the superior sensitivity of RMSMD to a quality change over the other four metrics.

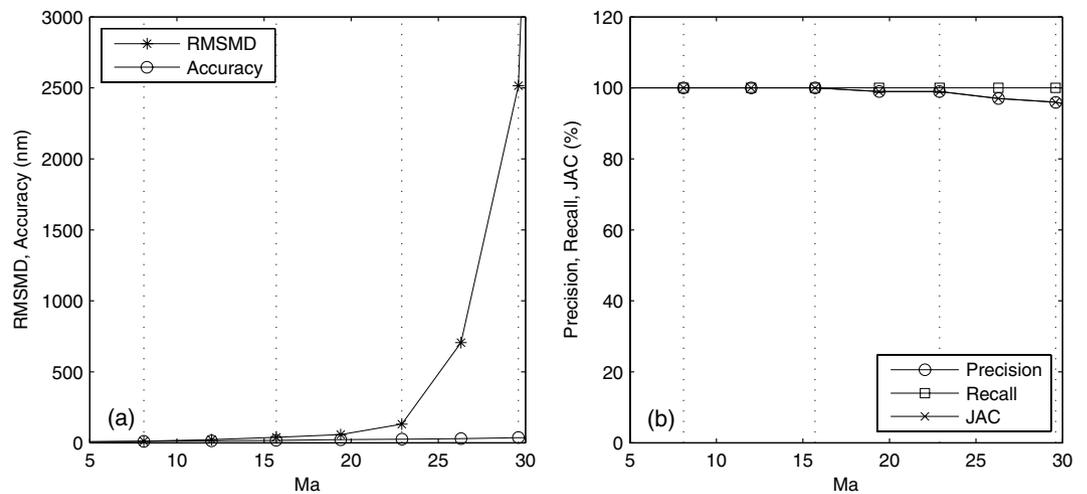

**Figure 5.** Metrics versus $M_a$, the mean number of activated emitters per frame. (**a**) RMSMD and accuracy, and (**b**) precision, recall, and JAC versus $M_a$. The vertical dotted lines from left to right correspond to $M_a$ in Fig. 4(a–d), respectively.

## Conclusions

We propose RMSMD as a quality metric for a localization nanoscopy image that consists of estimated emitter locations in comparison with their true locations. RMSMD depends only on the two sets of points regardless of how they are obtained. RMSMD measures how well two sets of points averagely, locally, and mutually fit to each other. As a distance metric, RMSMD possesses the properties common to distance metrics as well as its own unique properties. The four metrics of accuracy, precision, recall, and JAC inherently present ambiguity, discontinuity, and inappropriateness and fail to distinguish the qualities of different localization nanoscopy images in certain conditions. In contrast, RMSMD presents the advantages of universality, objectiveness, continuity, and sensitivity to a quality change over these four metrics. The unbiased Gaussian information-achieving estimator is a benchmark for practical localization algorithms. The RMSMD of an information-achieving localization nanoscopy image in addition to its visual quality benchmarks the quality of localization nanoscopy images. As a universal and objective metric, RMSMD can be broadly employed to evaluate the quality of localization nanoscopy images and the performance of localization algorithms and in general to measure the mutual fitness of two sets of points as well.

**Code availability.** The customer code of Matlab Version 7.5.0.342 (R2007b) that produces the figures is available online as supplementary material.

## Additional Information

**Competing Interests:** The author declares no competing interests.

**Publisher's note:** Springer Nature remains neutral with regard to jurisdictional claims in published maps and institutional affiliations.